\documentclass[letterpaper,journal]{IEEEtran}
\usepackage[T1]{fontenc}
\usepackage{color}
\usepackage{xcolor}
\usepackage{babel}
\usepackage{amsmath,amssymb,amsfonts}
\usepackage{algorithmic}
\usepackage{algorithm}
\usepackage{tabularx}
\usepackage{array}      
\usepackage{float}
\usepackage{diagbox}
\usepackage{subcaption,caption,epsfig}
\usepackage{makecell} 
\usepackage{pifont} 
\usepackage{multirow}
\usepackage{textcomp}
\usepackage{stfloats}
\usepackage{subcaption} 
\usepackage{url}
\usepackage{svg}
\usepackage{verbatim}
\usepackage{graphicx}
\usepackage{cite}
\usepackage{footnote}
\usepackage[dvips]{hyperref}
\usepackage{booktabs} 
\usepackage{tabularx}
\usepackage[none]{hyphenat}
\begin{document}

\title{Enabling Small-Signal Stability Analysis\\of Black-Box Voltage Source Converters\\in Large-Scale Modern Power Systems}

\author{Luis A. Garcia-Reyes, \IEEEmembership{Graduate Student Member, IEEE}, Josep Arévalo-Soler, \IEEEmembership{Member, IEEE}, \\ Oriol Gomis-Bellmunt, \IEEEmembership{Fellow, IEEE}, Eduardo Prieto-Araujo, \IEEEmembership{Senior Member, IEEE}, Vinícius A. Lacerda, \\ Macarena Martin-Almenta, Edgar Nuño-Martínez, and  Javier Renedo, \IEEEmembership{Senior Member, IEEE}
\vspace{-0.67cm}
\thanks{L. A. Garcia-Reyes, J. Arévalo-Soler, O. Gomis-Bellmunt, E. Prieto-Araujo and V. A Lacerda are with Centre d’Innovació Tecnològica en Convertidors Estàtics i Accionaments, Departament d’Enginyeria Elèctrica, Universitat Politècnica de Catalunya (CITCEA-UPC), Barcelona, 08028 ESP. M. Martin-Almenta, E. Nuño-Martínez and J. Renedo are with Red Eléctrica - Redeia, Madrid, 28109 ESP. Corresponding author: luis.reyes@upc.edu.}
}



\maketitle
\vspace{-8mm}
\begin{abstract}
Modern power systems increasingly rely on power electronic converters, yet many of these devices are provided as black-box models, limiting the applicability of conventional small-signal analysis (SSA) tools. This work presents a unified multi-variable fitted state-space (SSA-FITSS) methodology that enables accurate small-signal modeling of black-box Voltage Source Converters (VSCs) using frequency-domain (FD) identification, adaptive pole-expansion, and reduced-order realization. The method includes an automated state-interpretation strategy that assigns fitted states to representative control-loop categories based on their dominant frequency ranges, providing an approximate but meaningful physical interpretation of the identified dynamics. This capability allows extensive modal analysis, including eigenvalue sensitivities and participation factors, in systems where internal converter details are unavailable. The methodology is validated on a grid-following (GFL) VSC and applied to the New England system, which contains multiple black-box converters operating in both GFL and grid-forming (GFM) modes. Results show that the SSA-FITSS models accurately reproduce converter and system dynamics, support full eigenvalue-based analysis, and reveal stability limits under varying synchronous generation and GFL penetration levels. The approach overcomes key limitations of existing identification-based techniques by enabling scalable, interpretable, and system-wide stability assessment.
\end{abstract}


\vspace{-5mm}
\section{Introduction}
\label{sec:introduction}

\IEEEPARstart{T}{he} increasing penetration of renewable energy sources in modern power systems has intensified system complexity, particularly in stability assessment and operational analysis. Technologies such as wind turbines, solar photovoltaic (PV) generators, energy storage systems (ESS), hydrogen-electrolyzer  loads or data-center loads, among others,  are coupled to the power system through power-electronics converters. Typically, their dynamic models are delivered to Transmission System Operators (TSOs) as proprietary black-box models. These models restrict access to internal control structures and dynamic behavior, limiting their use mainly to non-linear time-domain simulation. As a result, conventional small-signal stability techniques cannot be applied directly and, in many cases, require assumptions that compromise analytical accuracy and system reliability~\cite{paperWIREs, paperMarc_blackbox, paper_black-box_reasons, paper_identification_perspectives}.

Two small-signal analysis (SSA) methods are widely standardized for modern power system applications: eigenvalue-based state-space analysis \cite{paper_CCM1,paper_CCM2} and impedance-based techniques~\cite{paperWIREs, cigre_928, paper_SISO}. The former requires full access to the mathematical model, including control structures and parameter values, enabling complete mode identification through linearization procedures~\cite{stamp_citcea,paper_SSA,paper_Benjamin} and supporting stability assessment through modal sensitivities and participation factors~\cite{paper_PFs1,paper_PFs2}. The latter relies on frequency-domain (FD) scanning to characterize the open-loop impedance between interconnected subsystems, from which closed-loop stability can be evaluated using the Generalized Nyquist Criterion, passivity analysis, and related metrics~\cite{paper_siad, paper_nyquist, paper_BodePlots}. Although impedance-based methods do not require access to internal converter models, they provide only local information and do not reveal system-wide modal behavior. Conversely, SSA eigenvalue-based methods cannot be applied when only black-box models are available. As a result, both approaches face fundamental limitations in black-box dominated environments.

Several alternative methods have been proposed to address these challenges. In~\cite{paper_n4sid}, the Numerical Algorithm for Subspace State-Space System Identification (N4SID) is used to synthesize state-space models directly from input–output time signals of black-box converters, enabling eigenvalue-based analysis. However, the multi-variable excitation required to ensure observability may induce nonlinear behavior in power electronic converters, reducing identification accuracy when high excitation levels are applied. In refs.~\cite{paper_cifuentes} and~\cite{paper_zhu}, an impedance-based identification framework using vector fitting is employed to obtain state-space representations. Although pole-sensitivity results are validated against time-domain simulations, these methods do not include validation against theoretical models and do not fully characterize pole locations or provide a physical interpretation of the identified modes. The method in~\cite{pccf}, based on pole clustering and vector fitting~\cite{paper_gustavsen_identification}, improves pole localization and enables order reduction by eliminating redundant states~\cite{paper_abner}, but the additional clustered poles have not been evaluated from a sensitivity perspective, and its performance under input–output data constraints or in large-scale studies remains insufficiently assessed. Ref.~\cite{paper_isgt} introduces a refined combination of these impedance-based approaches and demonstrates improvements in pole-location accuracy and model-order reduction, facilitating integration into SSA frameworks. However, it does not provide a clear order-selection criterion, lacks a systematic state-interpretation strategy, operates under constrained input–output data, and does not address scalability or robustness in large-scale applications. The ELSA method~\cite{paper_elsa} synthesizes minimal-order state-space models from FD impedance responses, avoiding overfitting and improving pole–zero placement, although its scalability and suitability for sensitivity-driven stability assessment have not yet been demonstrated. Finally, refs.~\cite{paper_geometrical} and~\cite{paper_IMF} introduce dominant-mode identification methods based on hyperplane geometry and iterative matrix fitting, respectively. While these approaches achieve high accuracy, no evidence is provided regarding their applicability to converter-dominated systems or their robustness under converter-driven interactions. Overall, existing methods do not offer a general-purpose, scalable, and interpretable framework capable of reliably identifying dynamic modes from black-box models while ensuring appropriate order selection without spurious dynamics.

To address these challenges and limitations, this work introduces a unified and robust multi-variable fitted state-space methodology (SSA-FITSS) that enables accurate SSA in large-scale systems with high penetration of black-box converters. The method characterizes the converter transfer matrix through FD identification using voltage, current, and frequency perturbations at the converter Point of Connection (POC). It relies on a vector-fitting Single-Input Single-Output (SISO) rational approximation strategy combined with an adaptive pole-expansion mechanism that inserts additional poles in regions of high local error when the convergence criterion is not satisfied, ensuring appropriate model order while avoiding overfitting and improving pole localization. A subsequent model-order reduction stage retains minimal yet representative dynamics, and a state interpretability strategy automatically assigns physical meaning to the identified states based on the characteristic frequency ranges of standardized control loops in grid-following (GFL) and grid-forming (GFM) Voltage Source Converters (VSCs). The methodology integrates seamlessly into SSA frameworks~\cite{stamp_citcea}, enabling conventional eigenvalue-based stability assessment without requiring proprietary internal models, and is validated through pole-location studies and modal decomposition analyses on a representative VSC. Its scalability and accuracy are demonstrated using the IEEE New England test system under different converter types and operating modes. To the authors’ knowledge, no existing vector-fitting-based methodology incorporates this multi-variable SISO approach with adaptive order refinement, automated state interpretability, and large-scale validation.

The article is organized as follows. Section~II describes the multi-variable SSA-FITSS algorithm and its state interpretability strategy. Section~III validates the methodology using a GFL VSC. Section~IV presents a case study based on the IEEE New England system, where black-box models are used to perform sensitivity and participation factor analyses.

\vspace{-4mm}
\section{Multi-Variable Fitted State-Space}

This work presents the SSA-FITSS methodology, previously introduced in~\cite{paper_isgt}, and incorporates a multivariable FD scanning technique to identify the transfer matrix of a black-box VSC through electromagnetic transient (EMT) simulations. The procedure applies small-signal perturbations in voltage and frequency at the converter POC while preserving system linearity conditions. Perturbation magnitudes are typically restricted to 1--3\% of nominal values. This selection arises because, within the state-space Component Connection Method (CCM), the VSC $dq$ output currents and frequency $\omega$ are commonly used as inputs to other blocks in the small-signal model~\cite{paper_CCM1,stamp_citcea}. As a result, the following FD representation of the VSC is obtained:
\begin{equation} \label{eq:Tvsc}
\begin{bmatrix}
   \Delta I_q (s) \\
   \Delta I_d (s) \\
   \Delta \omega_{\mathrm{c}}(s)
\end{bmatrix}
=
\begin{bmatrix}
    Y_{qq} (s) & Y_{qd} (s)  & R_{q \mathrm{g}} (s)\\
    Y_{dq} (s) & Y_{dd} (s)  & R_{d \mathrm{g}} (s)\\
    R_{\mathrm{g} q} (s) & R_{\mathrm{g} d} (s)  & R_{\mathrm{g} \mathrm{g}}(s)
\end{bmatrix}
\begin{bmatrix}
   \Delta V_q (s) \\
   \Delta V_d (s) \\
   \Delta \omega_{\mathrm{g}}(s)
\end{bmatrix}.
\end{equation}

\begin{figure}[t!]
    \centering
    \includegraphics[width=0.84\columnwidth]{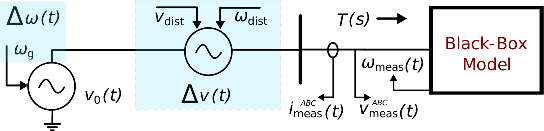}
    \caption{Scheme of the identification procedure based on~\cite{paper_siad}.}
    \label{fig:siad-tool}
    \vspace{-4.9mm}
\end{figure}

The \(3 \times 3\) matrix in~\eqref{eq:Tvsc}, denoted here and after as \(\mathbf{T}_{\text{VSC}}(s)\), represents the converter transfer matrix composed of admittance components and frequency-coupling terms in the \(dq\)-reference frame. The input frequency \(\Delta \omega_{\mathrm{g}}(s)\) corresponds to the global system frequency, while \(\Delta \omega_{\mathrm{c}}(s)\) denotes the perturbed internal frequency reference of the converter or black-box model. \(\Delta I_q(s)\), \(\Delta I_d(s)\), \(\Delta V_q(s)\), and \(\Delta V_d(s)\) represent the perturbed output currents and input voltages. The operator \(\Delta\) indicates increments from steady-state values, consistent with small-signal modeling principles~\cite{paperWIREs,stamp_citcea,smallsignalmolinas}. In (\ref{fig:siad-tool}), \(s = j\omega\) is the Laplace variable, where \(\omega\) is the angular frequency. When \(\Delta \omega_{\mathrm{c}}(s)\) is not accessible, the 2$\times$2 fitting approach is used~\cite{paper_isgt}. The open-source scanning tool used to obtain these FD responses is described in~\cite{paper_siad} and available in~\cite{siad-tool}. The perturbation scheme is shown in Fig.~\ref{fig:siad-tool}; a detailed description of the tool lies outside the scope of this work.

The transfer matrix is identified across a set of discrete frequency points, resulting in a three-dimensional dataset of dimensions \(J \times L \times f\), where \(J\) and \(L\) denote the number of system inputs and outputs, and \(f\) is the number of frequency samples. Although this study focuses on the \(dq0\) frame, the SSA-FITSS methodology assumes linear system behavior and is generalizable to other reference frames such as \(abc\) or \(pn0\).

To derive the state-space model, a rational approximation of \(\mathbf{T}_{\text{VSC}}(s)\) is first performed. Following the procedure in~\cite{paper_isgt}, each SISO entry of the transfer matrix is fitted independently using vector fitting~\cite{vectorfitting}. An initial pole set is defined by distributing poorly-damped complex-conjugate pairs across the frequency range of interest, adding a real pole only when an odd number of total poles is required. Since the true system order is unknown, the initial number of poles is intentionally chosen to be conservatively low. This avoids overparameterization at early stages while relying on the RMS convergence criterion and the adaptive pole-expansion mechanism to increase the order only when required. In this way, the algorithm identifies the minimal number of poles needed to achieve the prescribed accuracy. After convergence, the individual SISO models are assembled into a unified MIMO representation, and a subsequent order-reduction stage is applied to obtain a minimal state-space system~\cite{pccf,paper_gustavsen_identification,paper_abner,paper_isgt}.

\vspace{-4mm}
\subsection{SISO Fitting Approach} \label{sec:SISO_fit}

Each input–output subsystem of the transfer matrix in~\eqref{eq:Tvsc} is approximated independently using a rational pole–residue formulation. The fitting is performed with the \texttt{vecfit3} implementation of the vector fitting algorithm~\cite{SintefVectorFitting}. A conservative initial model order \(N\), here recommended between 4-8, is selected based on a preliminary inspection of the FD response, following the guidelines in~\cite{vectorfitting,pccf,paper_gustavsen_identification}. Since the adaptive pole-expansion mechanism automatically increases the order when necessary, the initial value of \(N\) does not need to be large. Instead, the algorithm converges toward the minimal order required to reproduce the dynamics with the prescribed accuracy. When \(N\) is even, all initial poles are assigned as complex-conjugate pairs. Their imaginary parts \(\beta_k\) are uniformly distributed over the frequency interval \([\min(f),\max(f)]\), and their real parts are set to small negative values proportional to \(\beta_k\). The resulting initial pole set is
\begin{equation}
\label{eq:initial_poles}
p_0 = -10^{-2}\beta_k \pm j\,\beta_k,
\qquad k = 1,2,\ldots, N/2.
\end{equation}
When \(N\) is odd, the same construction is applied to the complex-conjugate pairs, and one additional real pole is placed at the midpoint of the frequency range.

To ensure robust convergence, the fitting process incorporates an adaptive pole-expansion mechanism. After each iteration of \texttt{vecfit3}, the RMS fitting error between the measured response \(T_{\mathrm{VSC}}(s)\) and its approximation \(\hat{T}_{\mathrm{VSC}}(s)\) is evaluated. Convergence is achieved when \(\varepsilon_{\mathrm{RMS}}\) falls below a prescribed tolerance, typically \(10^{-6}\), which has been shown to provide very high accuracy for dynamic representation~\cite{paper_abner}. As shown in Algorithm~\ref{alg:redfit}, if the threshold is not met within a limited number of iterations, the algorithm enhances the pole set by inserting an additional pole in the frequency region exhibiting the largest local fitting error. The midpoint of this high-error interval is used as the nominal pole frequency: a real pole is added when the midpoint lies near DC, and a poorly-damped complex pole otherwise. The fitting procedure is then restarted with the expanded pole set.

\begin{algorithm}[t!]
\caption{Adaptive Pole-Expansion Strategy}
\label{alg:redfit}
\begin{algorithmic}[1]

\STATE \textbf{Input:} SISO response $T_{\mathrm{VSC}}(s)$, frequency samples $s_k$, initial poles $p_0$, tolerance $\varepsilon_{\mathrm{tol}}$
\STATE \textbf{Output:} Converged pole set $p^\star$

\STATE Initialize $p \gets p_0$, \; $\text{converged} \gets \text{false}$

\WHILE{not converged}

    \STATE Apply \texttt{vecfit3} using poles $p$ and obtain fitted response $\hat{T}_{\mathrm{VSC}}(s_k)$ and updated poles $\tilde{p}$

    \STATE Compute RMS error:
\[
        \varepsilon_{\mathrm{RMS}} =
        \sqrt{\frac{1}{f}\sum_{k=1}^{f} 
        \left|T_{\mathrm{VSC}}(s_k)-\hat{T}_{\mathrm{VSC}}(s_k)\right|^2}
    \]

    \IF{$\varepsilon_{\mathrm{RMS}} < \varepsilon_{\mathrm{tol}}$}
        \STATE $p^\star \gets \tilde{p}$, \; $\text{converged} \gets \text{true}$
        \STATE \textbf{break}
    \ENDIF

    \STATE Identify frequency of maximum local error: 
\[
        k^\star = \arg\max_k 
        \left|T_{\mathrm{VSC}}(s_k)-\hat{T}_{\mathrm{VSC}}(s_k)\right|
    \]

    \STATE Determine midpoint of the high-error interval:
\[
        s_{\mathrm{mid}} = \frac{s_{k^\star-1} + s_{k^\star+1}}{2}
    \]

    \IF{$|s_{\mathrm{mid}}| < 10^{-3}$}
        \STATE Insert real pole: $p_{\mathrm{new}} = \Re(s_{\mathrm{mid}})$
    \ELSE
        \STATE Insert poorly-damped complex pole:  
\[
            p_{\mathrm{new}} = -10^{-4}|s_{\mathrm{mid}}| 
            + j\,\Im(s_{\mathrm{mid}})
        \]

    \ENDIF

    \STATE Expand pole set: $p \gets [p; \; p_{\mathrm{new}}]$

\ENDWHILE

\STATE \textbf{return} $p^\star$
\vspace{-0.05cm}
\end{algorithmic}
\vspace{-0.05cm}
\end{algorithm}

After convergence, the resulting rational approximation of the transfer function is
\begin{equation}
\label{eq:Yvf}
T_{\mathrm{VSC}}(s)
\approx
\sum_{n=1}^{N} \frac{C_n}{s - p_n} + D + sE,
\end{equation}
where \(C_n\) are the complex residues, \(p_n\) are the fitted poles, \(D\) is a constant offset, and \(E\) accounts for the high-frequency asymptotic behavior. The upper frequency limit for fitting, \(f_{\mathrm{vecfit}}\), is selected according to the Nyquist criterion. In practice, the following rule is applied \cite{paper_isgt}:
\begin{equation}
\label{eq:rule_fs_vf}
f_{\mathrm{vecfit,max}} \leq \frac{f_s}{10},
\end{equation}
where \(f_s\) is the sampling frequency used during the transfer matrix identification. 

Once the rational model in~\eqref{eq:Yvf} is obtained, a structured state-space realization is constructed by assembling sub-blocks associated with each identified pole. In this formulation, the high-frequency term \(E\) is set to zero within \texttt{vecfit3} to enforce a strictly proper approximation, which simplifies the assembly of the realization. The global state-space matrices take the form:
\begin{subequations}
\begin{equation}\label{eq:ss_general1}
\mathbf{A} =
\begin{bmatrix}
\mathbf{A}_{p_1} & 0 & \cdots & 0 \\
0 & \mathbf{A}_{p_2} & \cdots & 0 \\
\vdots & \vdots & \ddots & \vdots \\
0 & 0 & \cdots & \mathbf{A}_{p_N}
\end{bmatrix},
\qquad
\mathbf{B} =
\begin{bmatrix}
\mathbf{B}_{p_1} \\
\mathbf{B}_{p_2} \\
\vdots \\
\mathbf{B}_{p_N}
\end{bmatrix},
\end{equation}
\begin{equation}\label{eq:ss_general2}
\mathbf{C} =
\begin{bmatrix}
\mathbf{C}_{p_1} & \mathbf{C}_{p_2} & \cdots & \mathbf{C}_{p_N}
\end{bmatrix},
\qquad
D = D.
\end{equation}
\end{subequations}

Each \(k\)-th pole contributes either a first-order or a second-order subsystem depending on whether it is real or part of a complex-conjugate pair:
\begin{itemize}
    \item Real poles \(p_k \in \mathbb{R}\):
    \begin{equation}\label{eq:block_real}
    \mathbf{A}_{p_k} = [\,p_k\,], 
    \qquad 
    \mathbf{B}_{p_k} = [\,1\,], 
    \qquad 
    \mathbf{C}_{p_k} = [\,C_k\,].
    \end{equation}

    \item Complex-conjugate poles \(p_k = \alpha \pm j\beta\):
    \begin{subequations}\label{eq:block_complex}
    \begin{equation}
    \mathbf{A}_{p_k} =
    \begin{bmatrix}
    \alpha & \beta \\
    -\beta & \alpha
    \end{bmatrix},
    \qquad
    \mathbf{B}_{p_k} =
    \begin{bmatrix}
    2 \\
    0
    \end{bmatrix},
    \end{equation}
    \begin{equation}
    \mathbf{C}_{p_k} =
    \begin{bmatrix}
    \mathrm{Re}(C_k) & \mathrm{Im}(C_k)
    \end{bmatrix}.
    \end{equation}
    \end{subequations}
\end{itemize}

\vspace{-0.5cm}
\subsection{Reduced-Order State-Space Modeling} \label{sec:redmet}

In large-scale SSA, reducing the dimensionality of state-space models is beneficial to improve computational efficiency and focus on the dominant dynamics relevant to system stability. In this work, the balanced realization (BR) method is selected for model reduction, since it provides comparable accuracy with lower complexity than singular value decomposition (SVD) which is evaluated in~\cite{paper_isgt}.

Starting from the state-space representation in~\eqref{eq:ss_general1} and~\eqref{eq:ss_general2},
\begin{equation}\label{ec38}
\begin{array}{l} 
\Delta \dot{\mathbf{x}} = \mathbf{A} \Delta \mathbf{x} + \mathbf{B} \Delta \mathbf{u}, \\
\Delta \mathbf{y} = \mathbf{C} \Delta \mathbf{x},
\end{array}
\end{equation}
the function \texttt{balreal}, detailed in \cite{balrealf}, is used to transform the system \((\mathbf{A}, \mathbf{B}, \mathbf{C})\) into a balanced form \((\mathbf{A}_b, \mathbf{B}_b, \mathbf{C}_b)\). In this representation, the controllability and observability Gramians are equal and diagonal, with entries corresponding to the square roots of the Hankel singular values.

To determine which states to retain, a compaction criterion based on the singular value ratio is applied~\cite{paper_gustavsen_identification}:
\begin{equation}\label{ec:tol}
    {\Sigma_r}/{\Sigma_1} < \sigma_{tol},
\end{equation}
where \(\sigma_{tol} > 5 \times 10^{-4}\) is the threshold used to identify negligible contributions. States associated with singular values below this threshold are removed using the \texttt{xelim} function, resulting in a reduced-order model \cite{paper_abner,paper_isgt}.

\vspace{-0.45cm}
\subsection{General Fitted State-Space Model} \label{sec:general_fss}

Once all SISO components of the transfer matrix in~\eqref{eq:Tvsc} have been individually fitted and reduced, the complete multi-variable state-space model is constructed by aggregating the local models. For MIMO systems, the overall structure is
{\normalsize
\begin{subequations}\label{eq:MIMO_ss}
\begin{align}
\mathbf{A} &= \text{diag}(\mathbf{A}_1, \mathbf{A}_2, \dots, \mathbf{A}_M), \mbox{\space \space}
\mathbf{B} &= 
\begin{bmatrix}
\mathbf{B}_{1,1} & \cdots & \mathbf{B}_{1,J} \\
\vdots & \ddots & \vdots \\
\mathbf{B}_{L,1} & \cdots & \mathbf{B}_{L,J}
\end{bmatrix}, \label{eq:MIMO_AB} \\
\mathbf{C} &= 
\begin{bmatrix}
\mathbf{C}_{1,1} & \cdots & \mathbf{C}_{1,L} \\
\vdots & \ddots & \vdots \\
\mathbf{C}_{J,1} & \cdots & \mathbf{C}_{J,L}
\end{bmatrix}, \mbox{\space \space}
\mathbf{D} &= 
\begin{bmatrix}
D_{1,1} & \cdots & D_{1,J} \\
\vdots & \ddots & \vdots \\
D_{L,1} & \cdots & D_{L,J}
\end{bmatrix}. \label{eq:MIMO_CD}
\end{align}
\end{subequations}
}
\color{black}

In this formulation, \(M\) represents the total number of fitted subsystems, each contributing a local state-space block. Depending on the specific input–output configuration, certain entries matrices $\mathbf{B}$ and $\mathbf{C}$ may be zero, reflecting decoupled or inactive dynamic paths. The resulting MIMO state-space model can then be integrated into the CCM framework for small-signal analysis~\cite{stamp_citcea}. This modular approach is designed to be broadly applicable to a wide range of black-box systems, regardless of their internal control architecture or complexity.

\vspace{-0.4cm}
\subsection{SSA-FITSS State Interpretability Strategy}

Identification methods such as SSA‑FITSS and those reviewed in~\cite{paper_n4sid,paper_cifuentes,pccf,paper_gustavsen_identification,paper_abner,paper_elsa,paper_geometrical,paper_IMF} produce abstract states \(x_1,x_2,\dots\) that lack direct physical correspondence to internal control loops or measurable variables. This differs from first‑principles small‑signal models, where each state maps to a specific physical or control quantity \cite{paper_PFs2}.

\begin{table}[t!]
\centering
\caption{Control-loop bandwidths of GFL and GFM VSCs relevant to oscillatory dynamics, with $\sigma \neq 0$ and $\omega \neq 0$.}
\label{tab:control-bands}
\vspace{-1mm}
\begin{tabular}{lccc}
\toprule
\textbf{Control Loop} &
\textbf{Bandwidth} &
\makecell[c]{\textbf{SSA-FITSS} \\ \textbf{Variable}} &
\textbf{Literature} \\
\midrule
Current & 0.1--2 kHz 
& $x_{cc}$ & \cite{TeodorescuBook,paperOnur} \\

Active-power / DC voltage & 1--100 Hz 
& $x_{vc}$ & \cite{TeodorescuBook,GuerreroDroop} \\

Reactive-power / AC voltage & 1--100 Hz 
& $x_{vc}$ & \cite{TeodorescuBook,GuerreroDroop} \\

PLL synchronization (sync) & 5--20 Hz 
& $x_{pll}$ & \cite{RodriguezPLL,FreijedoPLL} \\

P--f / Q--v droop control & 0.1--5 Hz 
& $x_{dp}$ & \cite{GuerreroDroop,PogakuMicrogrid} \\

Virtual inertia (GFM) & 0.1--5 Hz 
& $x_{vi}$ & \cite{ZhongBook,JiangGFM} \\

System electrical states & 40--80 Hz 
& $x_{sys}$ & \cite{TeodorescuBook,paperOnur} \\
\bottomrule
\end{tabular}
\vspace{-1mm}
\end{table}

\begin{table}[t!]
\centering
\caption{Bandwidths of GFL and GFM VSCs relevant to non-oscillatory dynamics, with $\sigma \neq 0$ and $\omega = 0$.}
\label{tab:control-bands_non_oscillatory}
\vspace{-1mm}
\resizebox{\columnwidth}{!}{%
  \begin{tabular}{lccc}
    \toprule
    \textbf{Type of dynamics} & \textbf{Bandwidth} & \makecell[c]{\textbf{SSA-FITSS} \\ \textbf{Variable}} & \textbf{Literature} \\
    \midrule
    Very fast (inner electrical) & $< -500\ \mathrm{s}^{-1}$ & $x_{n,il}$ & \multirow{5}{*}{\cite{TeodorescuBook,Hatziargyriou2021}} \\
    Fast (PQ, AC/DC control)     & $-20$ to $-500\ \mathrm{s}^{-1}$ & $x_{n,vc}$ &  \\
    Intermediate (sync)         & $-2$ to $-20\ \mathrm{s}^{-1}$   & $x_{n,syn}$ &  \\
    Slow (primary control)      & $-2$ to $-1\ \mathrm{s}^{-1}$    & $x_{n,pc}$ &  \\
    Very slow (other controls)  & $> -1\ \mathrm{s}^{-1}$         & $x_{n,oc}$ &  \\
    \bottomrule
  \end{tabular}%
}
\vspace{-0.3cm}
\end{table}

To address the lack of explicit physical meaning in the identified states, SSA‑FITSS adds an automated interpretation layer that maps fitted states to representative control‑loop categories using modal frequency, damping and participation. Modal data are obtained from the Jordan decomposition:
\begin{equation}\label{eq:decomp}
\mathbf{A} = \mathbf{R}\,\boldsymbol{\Lambda}\,\mathbf{R}^{-1},
\end{equation}
where \(\boldsymbol{\Lambda}\) is block‑diagonal with eigenvalues \(\lambda_i=\sigma_i\pm j\omega_i\) and \(\mathbf{R}\) contains right eigenvectors. Using the modal frequencies and participation from~\eqref{eq:decomp}, Table~\ref{tab:control-bands} summarizes representative states bandwidths for the most widely adopted control loops in VSCs, together with the symbolic labels used by SSA-FITSS to classify them. In addition to the oscillatory modes, Table~\ref{tab:control-bands_non_oscillatory} provides a complementary classification for non-oscillatory states, organized according to the characteristic scales of VSC dynamics. Finally, system electrical states typically appear near the fundamental frequency and arise from resonances or couplings among the converter, its filters, and the grid. 

This procedure yields an interpretable assignment of abstract fitted states to control dynamics while preserving the identification results. The assignment is heuristic and therefore applied only after the black‑box model has been identified as a GFL or GFM VSC. The resulting labels facilitate system‑level interpretation and enable consistent aggregation of states that share the same dynamic band.

\vspace{-3mm}
\subsection{Full SSA-FITSS Model: Black- and White-Box Integration}

Once SSA‑FITSS models have been generated and their states assigned, they are integrated into a full SSA‑FITSS-based representation of the complete modern power system. The formulation combines analytically derived, explicitly interpretable ``white‑box'' ($wb$) models, black‑box ($bb$) models, and intermediate ``gray‑box'' models \cite{paperMarc_blackbox}. Subsystems interconnect via their input–output relations, yielding the global SSA-FITSS-based representation:
\begin{equation}\label{eq:ss_general}
\begin{aligned} 
\Delta \dot{\mathbf{x}}_{f} &= \mathbf{A}_{f}\,\Delta \mathbf{x}_{f} + \mathbf{B}_{f}\,\Delta \mathbf{u}_{f}, \\
\Delta \mathbf{y}_{f} &= \mathbf{C}_{f}\,\Delta \mathbf{x}_{f} + \mathbf{D}_{f}\,\Delta \mathbf{u}_{f},
\end{aligned}
\end{equation}
with the partitioned state matrix and state vector
\begin{equation}\label{eq:partition_bw}
\mathbf{A}_{f} =
\begin{bmatrix}
 \mathbf{A}_{wb,wb} & \mathbf{A}_{wb,bb} \\
 \mathbf{A}_{bb,wb} & \mathbf{A}_{bb,bb}
\end{bmatrix},
\qquad
\mathbf{x}_{f} =
\begin{bmatrix}
\mathbf{x}_{wb} \\
\mathbf{x}_{bb}
\end{bmatrix},
\end{equation}
where $\mathbf{A}_{wb,wb}$ and $\mathbf{A}_{bb,bb}$ denote the pure white‑box and black‑box dynamics, respectively, and $\mathbf{A}_{wb,bb}$, $\mathbf{A}_{bb,wb}$ are the cross-coupling blocks. For each eigenvalue $\lambda_j$ of $\mathbf{A}_{f}$, the participation factor of state $k$ in mode $j$ is given by \cite{paper_PFs2}:
\begin{equation}\label{ec:pfs}
    p_{kj} = \boldsymbol{\ell}_{jk}\, \mathbf{r}_{kj},
\end{equation}
where $\boldsymbol{\ell}_{j}$ and $\mathbf{r}_{j}$ are the left and right eigenvectors of mode $j$ and $\ell_{jk}$, $r_{kj}$ denote their $k$-th components. The total contribution of black‑ and white‑box states to mode $j$ is then:
\begin{subequations}\label{eq:PF_full}
\begin{equation}
P_j^{({bb})}=\sum_{k\in\mathcal{K}_{bb}} \left|p_{kj}\right|,
\end{equation}
\begin{equation}
P_j^{({wb})}=\sum_{k\in\mathcal{K}_{wb}} \left|p_{kj}\right|,
\end{equation}
\end{subequations}
where $\mathcal{K}_{bb}$ and $\mathcal{K}_{wb}$ are the index sets of black‑box and white‑box states. Finally, the modes are classified as:
\begin{itemize}
\item {Black‑box dominated} if $P_j^{({bb})}\gg P_j^{({wb})}$.
\item {White‑box dominated} if $P_j^{({wb})}\gg P_j^{({bb})}$.
\item {Hybrid (coupled)} if $P_j^{({bb})}$ and $P_j^{({wb})}$ are comparable.
\end{itemize}

\vspace{-3mm}

\section{Multi-Variable SSA-FITSS Method Validation} \label{sec:GFLvalidation}

\begin{figure}[t!]
    \centering
    \includegraphics[width=0.8\columnwidth]{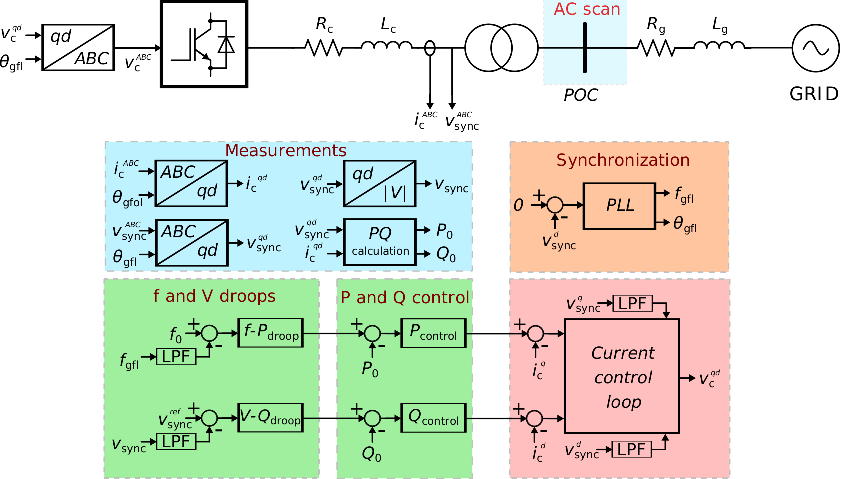}
    \vspace{-1mm}
    \caption{GFL control scheme.}
    \label{fig:gfl_scanner}
    \vspace{-2.5mm}
\end{figure}

To evaluate the performance of the extended SSA-FITSS methodology in a multi-variable setting, a GFL VSC is selected as the test case. The system configuration is shown in Fig.~\ref{fig:gfl_scanner}, and its control architecture is described in~\cite{ss_citcea2}. The active and reactive power control loops use proportional–integral (PI) controllers, while the frequency–power ($P$–$f$) and voltage–reactive power ($Q$–$V$) droop mechanisms are implemented with proportional control. The converter and control parameters used in the validation are listed in Table~\ref{table:gfl_parameters}.

\begin{table}[t!]
\caption{GFL converter parameters\label{table:gfl_parameters}}
\vspace{-1mm}
\centering
\begin{tabular}{lcc|lcc}
\toprule
\textbf{Parameter} & \textbf{Value} & \textbf{Unit} & \textbf{Parameter} & \textbf{Value} & \textbf{Unit} \\
\midrule
$S_{base}$  & 2.750 & MVA   & $k_{i_{PLL}}$ & 568.1 & pu    \\  
$V_{base}$  & 690.0 & V     & $k_{p_{PLL}}$ & 1.420 & pu    \\  
$SCR$        & 3.000     & -     & $k_{i_{CCL}}$ & 0.866 & pu    \\  
$X/R$         & 10.00    & -     & $k_{p_{CCL}}$ & 1.378 & pu    \\  
$V_{DC}$    & 2070  & V     & $k_{i_{PQ}}$  & 1.000 & pu    \\  
$R_{C}$     & 0.034 & $\Omega$ & $k_{p_{PQ}}$ & 0.100 & pu    \\  
$L_{C}$     & 43.00 & mH    & $\tau_{PQ}$  & 50.00 & ms    \\  
$R_{trans}$ & 0.003 & $\Omega$ & $k_{P}$     & 0.175 & pu/pu    \\  
$L_{trans}$ & 4.300 & mH    & $k_{Q}$      & 0.073 & pu/pu    \\  
\bottomrule
\end{tabular}
\vspace{-5mm}
\end{table}

The transfer matrix is identified using the open-source SIaD-Tool in~\cite{siad-tool}, with the converter operating at \(P = 0.5\ \text{pu}\) and \(Q = 0.1\ \text{pu}\). A balanced operating condition is assumed, and a single-tone identification scheme is applied using 110 linearly spaced frequency points from 0.1~Hz to 1~kHz. The perturbation consists of a 1\% series voltage injection relative to \(V_{base}\) and a 0.05\% frequency deviation around 50~Hz. A detailed description of the scanning tool is provided in~\cite{paper_siad}.

For the SSA-FITSS model synthesis, an initial set of eight complex-conjugate poles is generated using~\eqref{eq:initial_poles} across the frequency range from 1~mHz to 1~kHz. The fitting is performed using a $\varepsilon_{\mathrm{RMS}}$ tolerance of \(5\times10^{-4}\). After applying the full multi-variable SSA‑FITSS methodology, the resulting model is stable and of order 16, with 4 states classified as \(x_{n,\mathrm{syn}}\), 6 as \(x_{n,\mathrm{vc}}\), 4 as \(x_{n,\mathrm{il}}\), and 2 as \(x_{\mathrm{vc}}/x_{\mathrm{sys}}\), according to the automated state‑interpretation strategy. Note that variables sharing the same band are accommodated together to improve dynamic assignment. The complete fitting process requires only three pole-expansion iterations and executes in 0.719~s using MATLAB R2024b on a 1.6~GHz Intel i5 16~GB RAM.

\begin{figure}[t!]
    \centering
    \includegraphics[width=0.985\columnwidth]{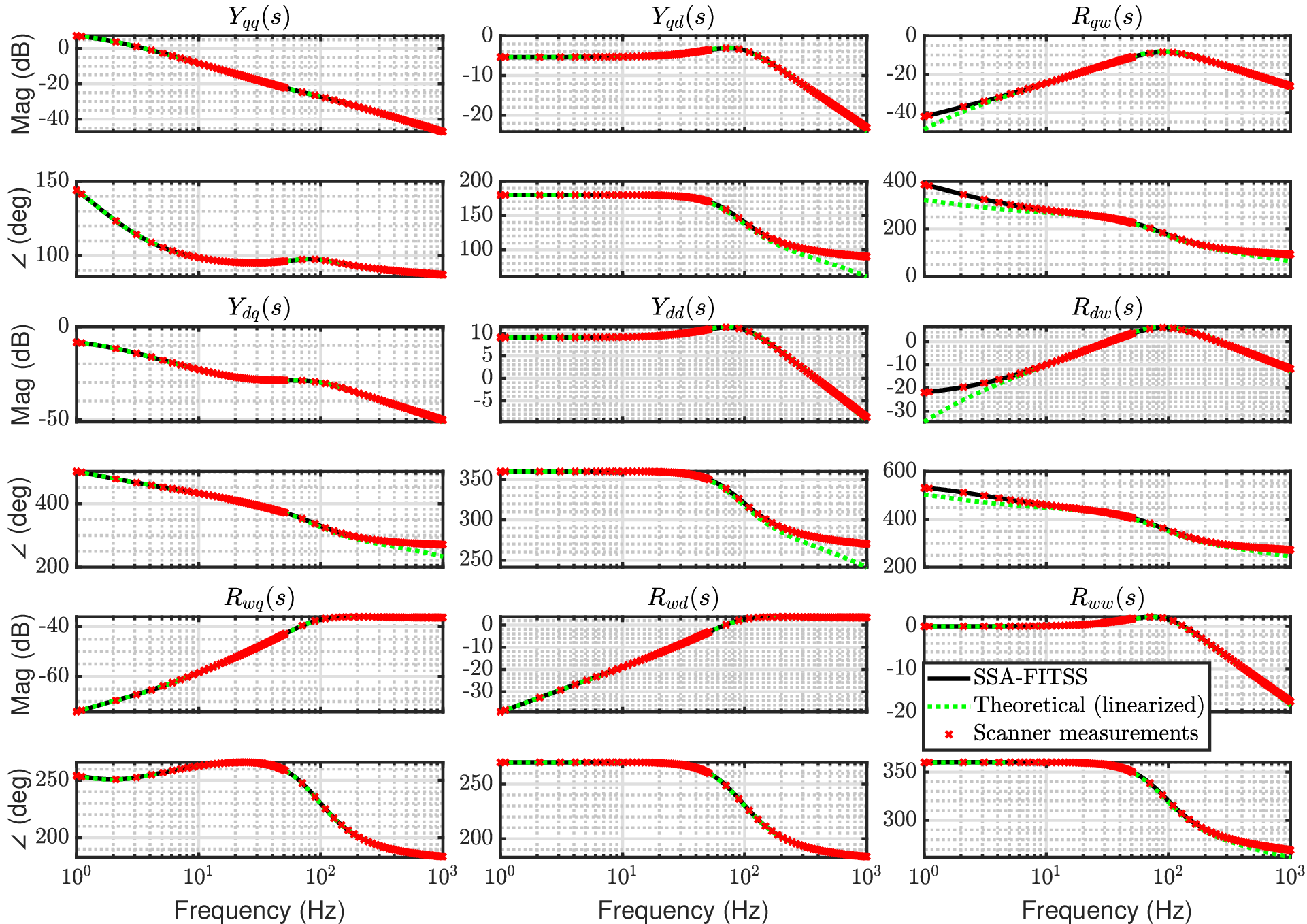} \vspace{-1mm}
    \caption{Frequency response of the SSA-FITSS model.}
    \label{fig:GFL-fitted}
    \vspace{-0.67cm}
\end{figure}

Validation is performed by comparing the FD response of the SSA-FITSS-based model with that of a linearized state-space representation of the GFL VSC developed following~\cite{stamp_citcea,ss_citcea2}. The input signals used for comparison are \(\Delta V^{POC}_{dq}\) and \(\Delta \omega^{POC}_{g}\), and the outputs are \(\Delta I^{POC}_{dq}\) and \(\Delta \omega^{gfl}_{c}\). As shown in Fig.~\ref{fig:GFL-fitted}, the SSA-FITSS model exhibits excellent agreement with both the theoretical response and the FD measurements, accurately capturing all coupling terms in the transfer matrix. Minor deviations at very low frequencies in the $R_{q\omega}(s)$ and $R_{d\omega}(s)$ elements arise from the perturbation strategy and PLL interaction, which complicate matching the theoretical response but still yield an accurate representation.

\begin{figure}[t!]
    \centering
    \begin{subfigure}[b]{\columnwidth}
        \centering
        \includegraphics[width=0.8\textwidth]{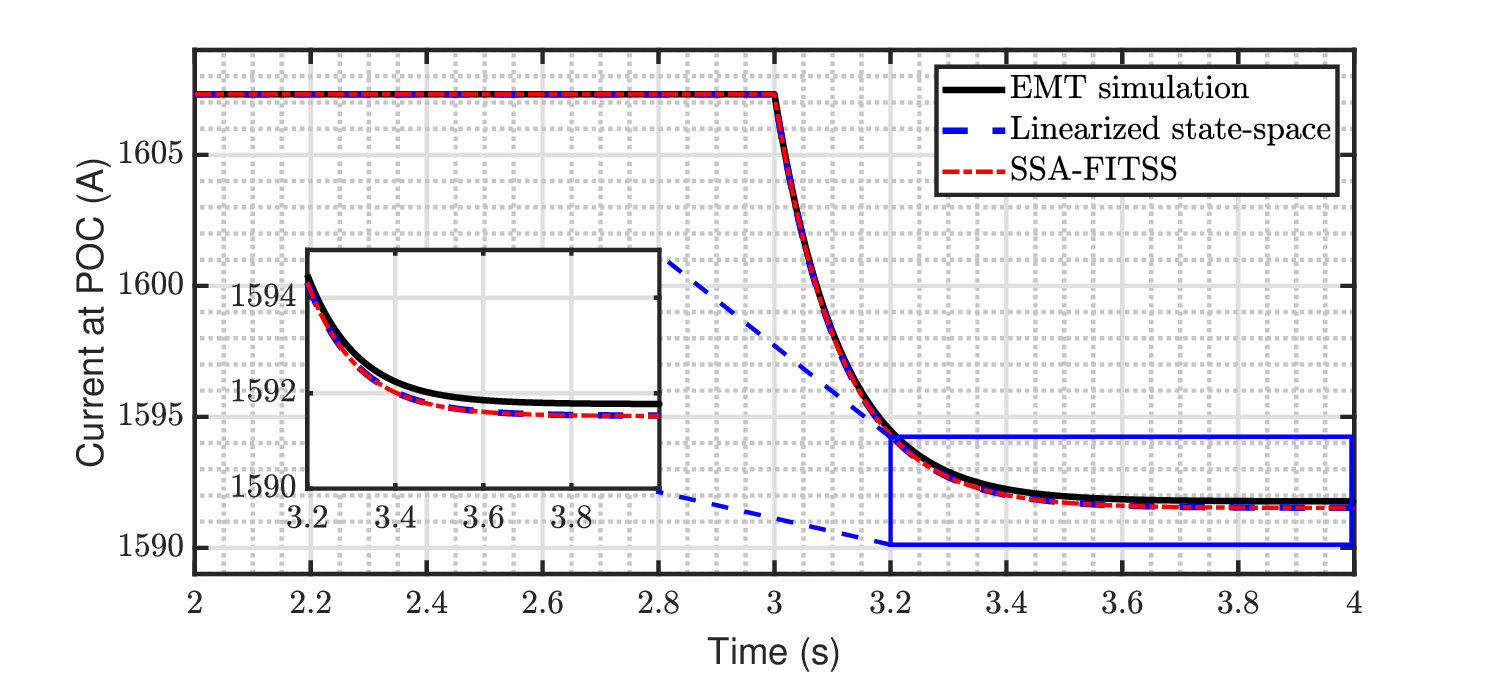}
        \vspace{-2mm}
        \caption{Instantaneous current magnitude at POC.}
        \label{fig:current_rms}
    \end{subfigure}
    \begin{subfigure}[b]{\columnwidth}
        \centering
        \includegraphics[width=0.8\textwidth]{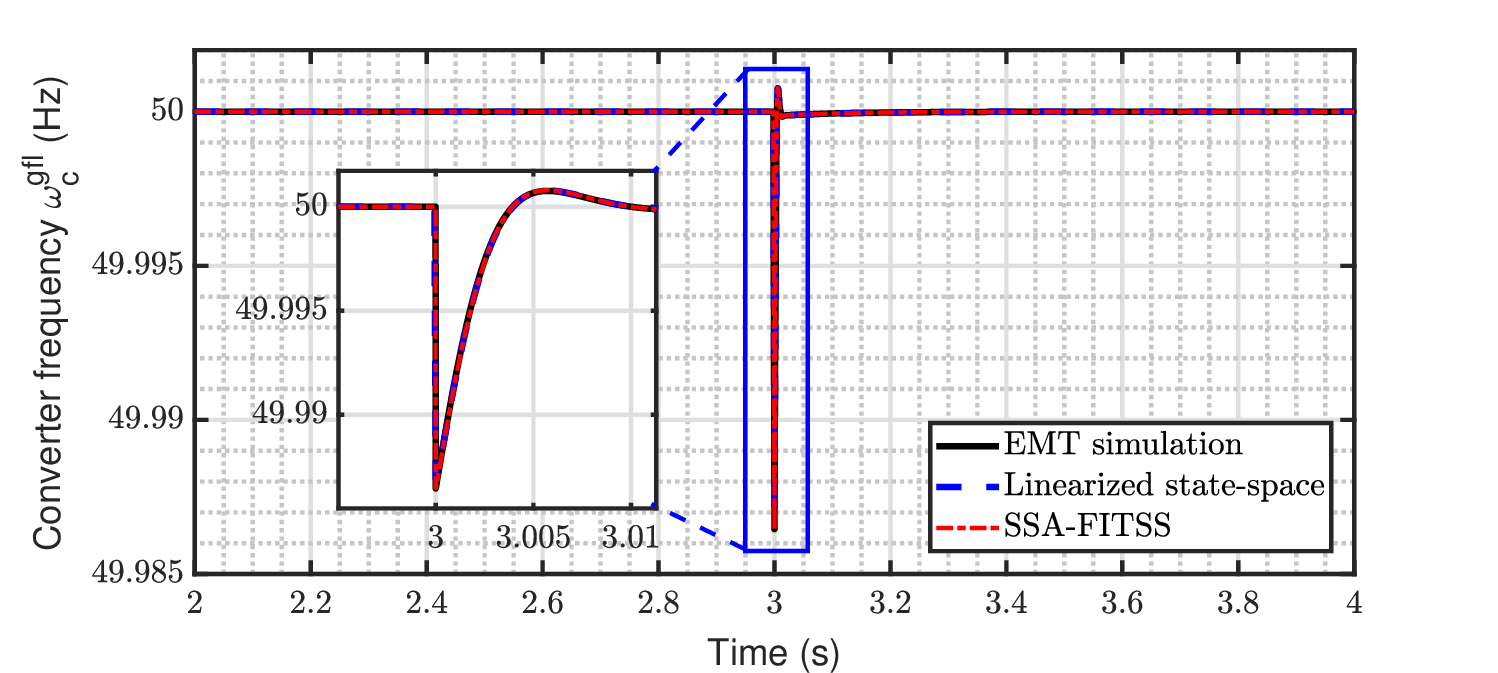}
        \vspace{-2mm}
        \caption{Converter frequency from $\omega^{gfl}_{c}$.}
        \label{fig:w_pll}
    \end{subfigure}
    \vspace{-5mm}
    \caption{Validation of SSA-FITSS for the GFL VSC.}
    \label{fig_1}
    \vspace{-4mm}
\end{figure}

To further assess the fidelity of the SSA-FITSS model, time-domain simulations are performed under small-signal perturbations. Fig.~\ref{fig_1} shows the system response to a 1\% step disturbance in the nominal voltage \(V_{base}\), comparing EMT simulations, the linearized model, and the SSA-FITSS-based representation. The results confirm that the SSA-FITSS methodology captures the converter dynamic behavior with high accuracy. The nadir of the response is reproduced with precision, demonstrating the capability of SSA-FITSS to extract high-fidelity dynamics.

\vspace{-4mm}
\subsection{Pole Location and Mode Analysis}

\begin{figure}[t!]
    \centering
    \begin{subfigure}[b]{0.8\columnwidth}
        \centering
        \includegraphics[width=\textwidth]{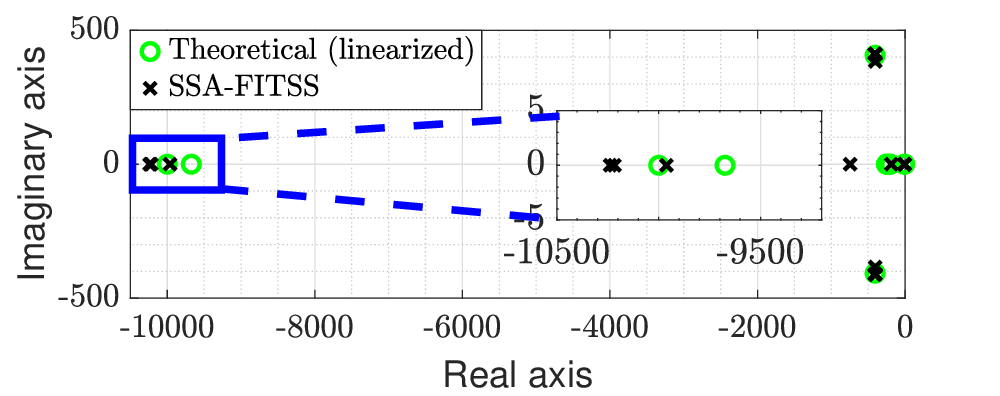}
        \vspace{-6mm}
        \caption{Fast-damping region.}
        \label{fig:poles_GFLa}
    \end{subfigure}
    \begin{subfigure}[b]{0.8\columnwidth}
        \centering
        \includegraphics[width=\textwidth]{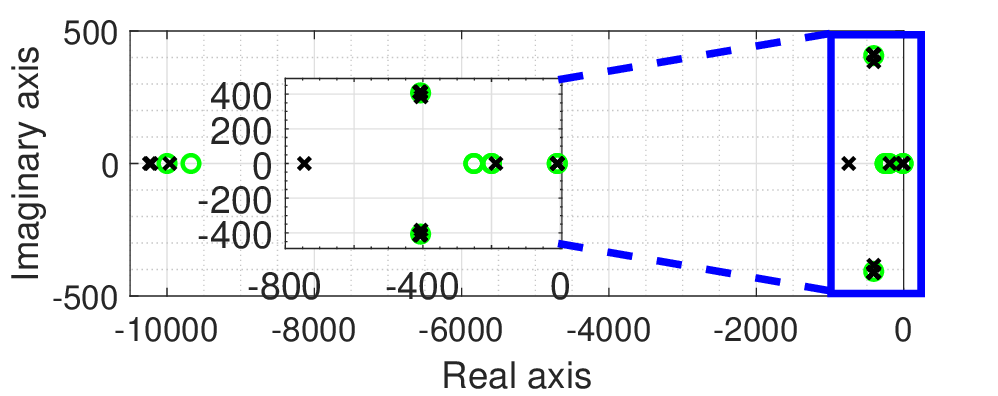}
        \vspace{-6mm}
        \caption{Slow-damping region.}
        \label{fig:poles_GFLb}
    \end{subfigure}
    \vspace{-1mm}
\caption{Pole-location comparison for the GFL VSC.}
\label{fig:poles_GFL}
\vspace{-6mm}
\end{figure}

The theoretical model exhibits a system order of 10, whereas the SSA-FITSS-based model results in 16 poles, 10 real and 3 complex-conjugate pairs. This increase does not indicate additional dominant dynamics; it arises from the pole-expansion strategy, which introduces poles in regions of high RMS fitting error. As shown in Fig.~\ref{fig:poles_GFL}, the theoretical mode at 64.8~Hz is represented in the SSA-FITSS model by 3 closely spaced poles, two near 65.3~Hz and one near 65.4~Hz.

Table~\ref{table:poles} shows the numerical summary of the identified poles, the obtained errors, and the variables assigned by the algorithm. The average pole‑identification error equals 1.53\% for the applied order‑reduction strategy. As illustrated in Fig.~\ref{fig:poles_GFL}, the SSA‑FITSS model captures modes in both fast‑ and slow‑damping regions with minor deviations. Two observations arise: the theoretical mode at \(-251\)~rad/s is not captured, and an additional mode appears near \(-744\)~rad/s. These discrepancies are examined in the next subsection.

Following the classification in Table~\ref{tab:control-bands}, the modes at 65.3~Hz and 65.4~Hz correspond to the supersynchronous grid‑interaction category \(x_{vc}\), \(x_{sys}\), reflecting coupling between the converter and the grid. The remaining non‑oscillatory modes are associated with synchronization dynamics \(x_{n,syn}\), voltage and power controllers \(x_{n,vc}\), and very fast inner‑electrical‑loop dynamics \(x_{n,il}\). This assignment is automatically generated by the SSA-FITSS algorithm. Overall, as shown in Table~\ref{table:poles}, SSA‑FITSS extracts the critical modes with minor deviations: additional poles appear but are close enough to be considered practically identical.

\begin{table}[!t]
\caption{Pole-location results using SSA-FITSS\label{table:poles}}
\vspace{-1.5mm}
\centering
\begin{tabular}{cccccc|}
\toprule
\textbf{Theoretical} & \textbf{Fitted with} & \textbf{Error} & \textbf{SSA-FITSS}\\ 
\textbf{poles} & \textbf{SSA-FITSS} & \textbf{(\%)} & \textbf{variable}\\ 
\midrule
-8.555 + j0.000&-8.411 + j0.000& 1.68 & $x_{n,syn}$\\

-&-8.542 + j0.000& - & $x_{n,syn}$\\

-8.655 + j0.000&-8.649 + j0.000& 0.07 & $x_{n,syn}$ \\

-&-8.651 + j0.000& - & $x_{n,syn}$\\

-199.9 + j0.000&197.1 + j0.000& 1.40 & $x_{n,vc}$\\

-200.0 + j0.000&197.1+ j0.000& 1.45 & $x_{n,vc}$\\

-251.3 + j0.000& - & - & -\\

-251.3 + j0.000&- &- & - \\

-406.5 - j407.1&-406.6 - j410.8& 0.64 & $x_{vc}$/$x_{sys}$\\

-406.5 + j407.1&-406.6 + j410.8& 0.64 & $x_{vc}$/$x_{sys}$\\

-&-406.6 - j410.8& - & $x_{vc}$/$x_{sys}$\\

-&-406.6 + j410.8& - & $x_{vc}$/$x_{sys}$\\

-&-406.5 - j411.1& - & $x_{vc}$/$x_{sys}$\\

-&-406.5 + j411.1& - & $x_{vc}$/$x_{sys}$\\

-&-744.6 + j0.000& - & $x_{n,il}$\\

-&-996.2 + j0.000& - & $x_{n,il}$\\

-9774 + j0.000&-10189 + j0.000& 4.25 & $x_{n,il}$\\

-10000 + j0.000&-10214 + j0.000& 2.14 & $x_{n,il}$\\
\bottomrule
\end{tabular}
\vspace{-4mm}
\end{table}

\vspace{-4.0mm}
\subsection{Modal Decomposition-Based Time Response Analysis}\label{sec:modaldecomposition}

Although the SSA-FITSS methodology proposed in this work yields high-fidelity dynamic models, it is necessary to evaluate the contribution of individual modes to specific input–output channels. This analysis determines whether modes not captured during fitting have a meaningful impact on the system dynamics. Since not all modes are equally observable or controllable from a given perturbation location, a modal decomposition-based time-domain analysis is used to quantify the influence of each mode in the theoretical system. Considering the LTI system:
\begin{equation}\label{ec38b}
\begin{array}{l} 
\dot{\mathbf{x}}(t) = \mathbf{A} \mathbf{x}(t) + \mathbf{B} \mathbf{u}(t), \\
\mathbf{y}(t) = \mathbf{C} \mathbf{x}(t) + \mathbf{D} \mathbf{u}(t),
\end{array}
\end{equation}
where the state matrix \(\mathbf{A}\) is diagonalized by eigenvalue decomposition using the procedure given in \eqref{eq:decomp}. Transforming the system into modal coordinates using~$\mathbf{R}$:
\begin{equation}
\mathbf{B}_{\text{modal}} = \mathbf{R}^{-1} \mathbf{B}, \qquad 
\mathbf{C}_{\text{modal}} = \mathbf{C} \mathbf{R}.
\end{equation}

The response of mode \(i\) to an unit step input is:
\begin{equation}
y_i(t) = \mathrm{Re}\!\left[ \frac{r_i}{\lambda_i} \left( e^{\lambda_i t} - 1 \right) \right],
\end{equation}
where the modal residue is defined as:
\begin{equation}
r_i = C_k^{(i)} B_i^{(j)},
\end{equation}
with \(C_k^{(i)}\) and \(B_i^{(j)}\) denoting the modal projections of the output and input matrices for output channel \(k\) and input channel \(j\). The magnitude of \(r_i\) indicates the degree of modal participation: large residues correspond to dominant dynamics, whereas small residues imply negligible influence.

\begin{figure}[t!]
\centering
\includegraphics[width=0.95\columnwidth]{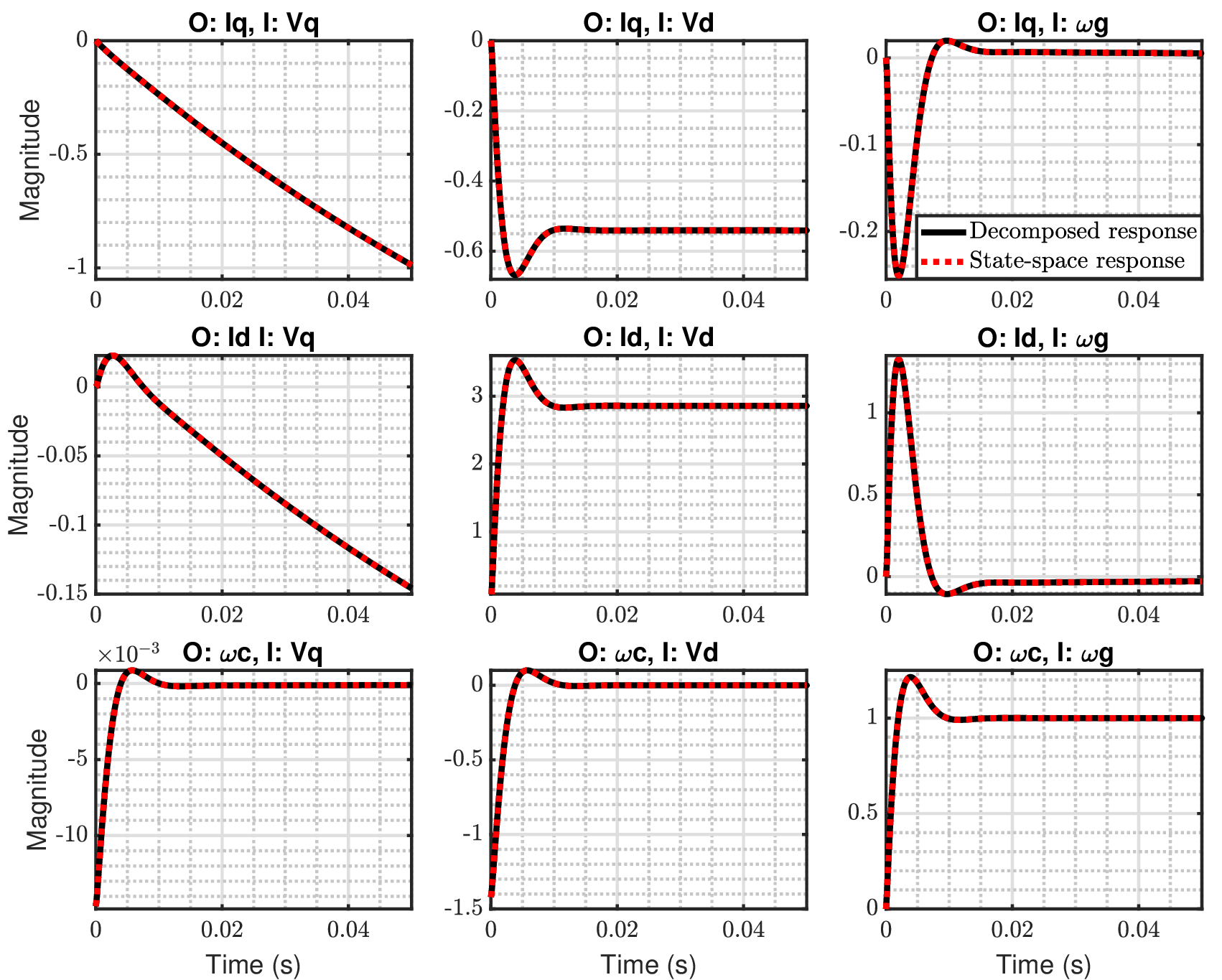}
\vspace{-1.5mm}
\caption{Transient responses for all input–output combinations from linear model under modal decomposition analysis.}
\label{fig:tdmodes}
\vspace{-6mm}
\end{figure}

This decomposition is used to assess the impact of the theoretical modes at \(-251\)~rad/s, which are not identified by the SSA-FITSS model. Fig.~\ref{fig:tdmodes} compares the transient responses obtained from the full theoretical model with those obtained after removing this mode. As shown, excluding the \(-251\) mode produces negligible deviations across all input–output channels, indicating that this mode has minimal dynamic relevance. A similar test was performed for the additional pole at \(-744\)~rad/s introduced by the fitting process. The results show only minor deviations in the frequency–grid and \(d\)-axis output coupling, without affecting the remaining responses.

These findings indicate that the modes not captured by the SSA-FITSS model do not significantly contribute to the system dynamics, and that the additional aggregated modes introduced during fitting have limited practical impact. To further assess modal relevance, a participation factor analysis is recommended to identify which modes meaningfully influence the full system dynamics~\cite{paper_PFs2}. 

\vspace{-2.5mm}
\section{Case Study: New England Power System}\label{sec:casestudy}

\begin{figure}[t!]
\centering
\includegraphics[width=0.98\columnwidth]{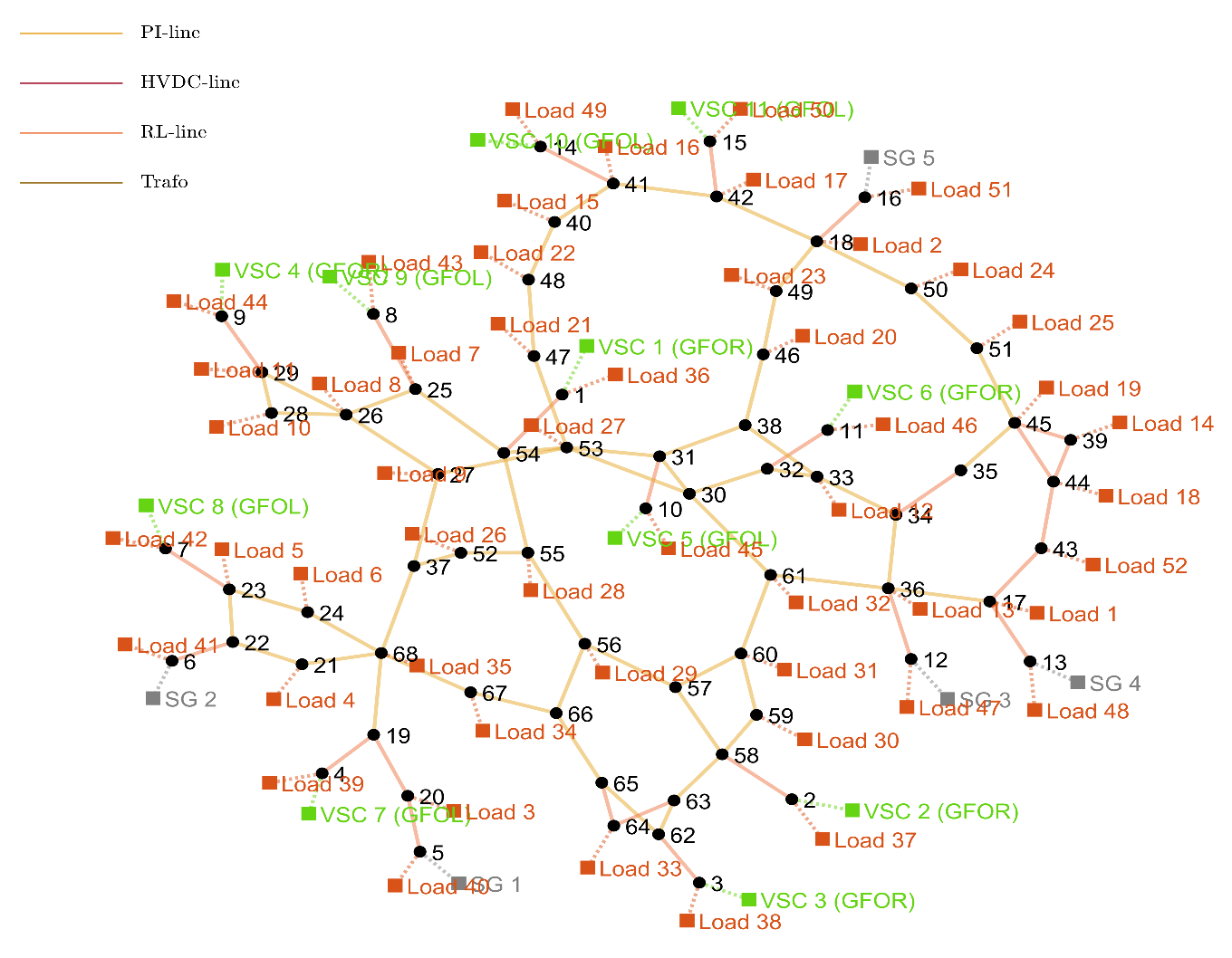}
\vspace{-2.5mm}
\caption{Layout of the New England power system.}
\label{fig:New_England}
\vspace{-5mm}
\end{figure}

To demonstrate the robustness, scalability, and practical applicability of the extended SSA-FITSS methodology, the New England power system is used as a benchmark. Detailed system parameters and configuration data are available in~\cite{stamp_citcea}. The 63-bus, 50~Hz network includes 11 VSCs, six operating in GFL mode and five in GFM mode, together with five synchronous generators (SGs) modeled with full dynamic detail. Among the VSCs, units 3, 4, 5, 7, 8, and 10 are treated as black‑box models for the SSA‑FITSS application, although full detail is available. In realistic settings, only limited information is accessible; thus, nominal power and voltage, input/output frequency, control mode, and selected control‑loop constants are assumed known. All VSCs operate under PQ control. Transmission lines are modeled using $\pi$-section equivalents, and loads are represented as concentrated $RL$ parameters. Fig.~\ref{fig:New_England} illustrates the system layout, highlighting the location of the VSC and SG units.

The FD hybrid admittance data for the black-box VSCs is obtained using the SIaD-Tool~\cite{siad-tool}, while the full linear white-box system is constructed using the STAMP tool~\cite{stamp_citcea}. For each black-box VSC and operating point, the SSA-FITSS algorithm is applied to synthesize a reduced-order state-space model, which is then integrated into STAMP to enable complete eigenvalue-based analysis of the entire network in an automated manner.

The case study is structured into three stages. First, a validation study is performed using the SSA-FITSS methodology to extract small-signal models from the black-box VSCs and evaluate their accuracy against the SSA-EMT theoretical model in a multi-variable, large-scale context. Second, a scenario is examined in which the rated capacity of the SGs is progressively reduced to emulate increasing converter penetration. Stability limits are evaluated through eigenvalue sensitivity analysis. Finally, a complementary scenario is considered in which the rated capacity of the GFM converters is reduced, resulting in a system dominated by GFL operation.

\vspace{-0.37cm}
\subsection{Black-Box Dominated System}\label{sec:black_box_validation}

The EMT frequency scan is performed using a time step of 10~$\mu$s, a frequency resolution of 1~Hz, and 150 linearly spaced frequency points. A voltage and frequency perturbation of 1\% and 0.05\% of the nominal 138~kV and 50 Hz are applied, with a system base power of 100~MVA. An initial order of 4 poles is selected within the frequency range of 1~mHz to 1~kHz for all black-box VSCs, using a $\varepsilon_{\mathrm{RMS}}$ tolerance of $5\times10^{-3}$. The resulting final orders for each VSC are summarized in Table~\ref{tab:final_orders}. The full system contains 824 states, of which 259 are from the black-box VSCs. All identified SSA-FITSS models are confirmed to be stable using MATLAB’s \texttt{isstable} function. Compared with the white‑box VSC models (average order 24 for GFL and 27 for GFM), the SSA‑FITSS models exhibit higher order due to the adaptive pole‑expansion strategy; however, as validated in Section~\ref{sec:GFLvalidation}, the resulting “over‑fitting” mainly introduces additional poles associated with existing dynamics rather than spurious modes.

\begin{table}[t!]
\centering
\caption{Results from the identified black-box models.}
\vspace{-1mm}
\label{tab:final_orders}
\begin{tabular}{ccccc}
\toprule
\textbf{Converter ID} & \textbf{Order} & \textbf{ID Time (s)} & Final $\mathbf{\epsilon_{RMS}}$ & \textbf{Status} \\
\midrule
GFM3  & 55 & 1.98 & 1.92$\times 10^{-3}$ & Stable \\ 
GFM4  & 51 & 1.63 & 1.89$\times 10^{-3}$ & Stable \\ 
GFL5  & 37 & 0.85 & 1.54$\times 10^{-3}$ & Stable \\ 
GFL7  & 41 & 0.90 & 1.72$\times 10^{-3}$ & Stable \\ 
GFL8  & 36 & 0.71 & 1.12$\times 10^{-3}$ & Stable \\ 
GFL10 & 39 & 1.03 & 1.43$\times 10^{-3}$ & Stable \\ 
\bottomrule
\end{tabular}
\vspace{-3.5mm}
\end{table}

\begin{figure}[t!]
    \centering
    \begin{subfigure}[b]{0.75\columnwidth}
        \centering
        \includegraphics[width=\textwidth]{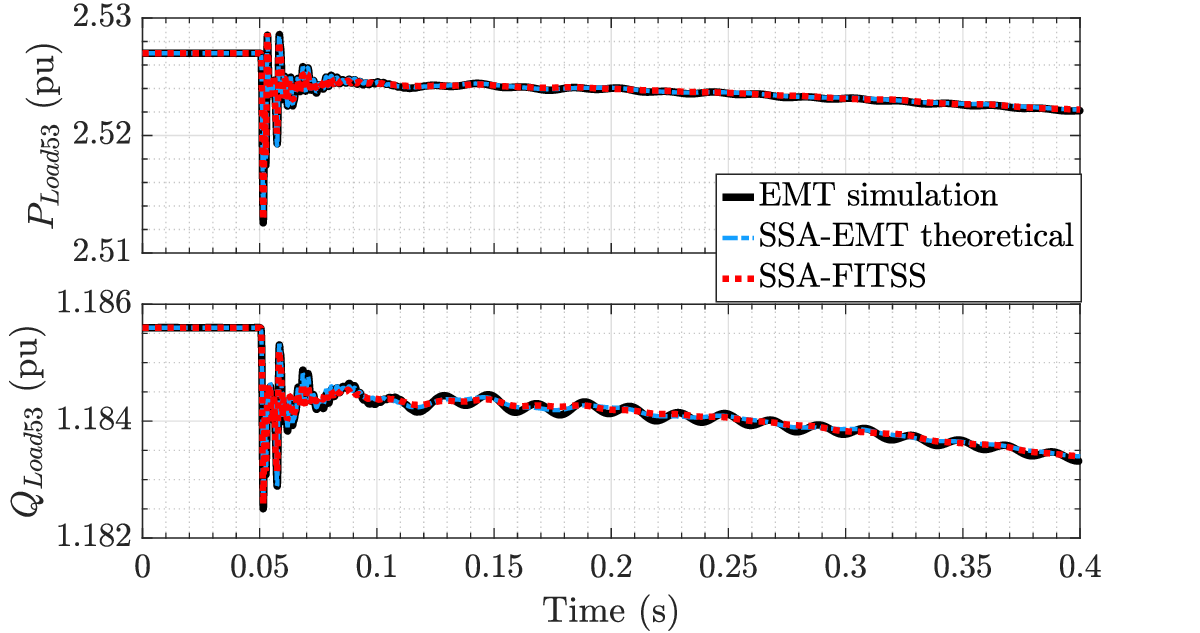}
        \vspace{-6mm}
        \caption{Power consumed by load 53.}
        \label{fig:V6}
    \end{subfigure}
    \begin{subfigure}[b]{0.75\columnwidth}
        \centering
        \includegraphics[width=\textwidth]{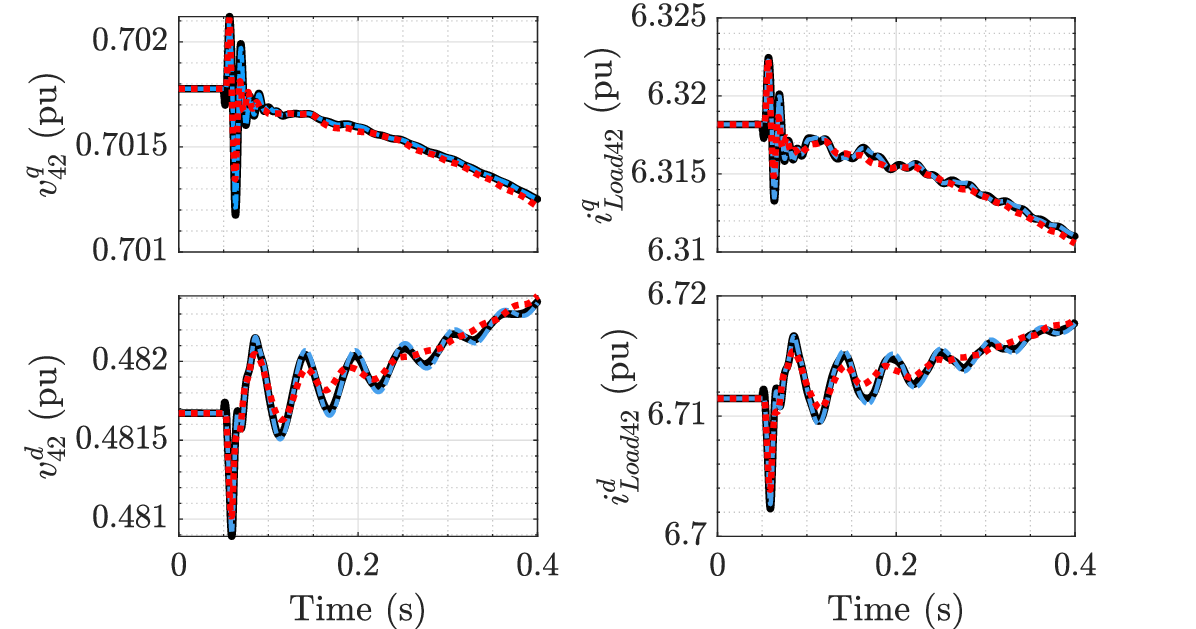}
        \vspace{-6mm}
        \caption{Voltages and currents at bus 42.}
        \label{fig:PQ2}
    \end{subfigure}
    \begin{subfigure}[b]{0.75\columnwidth}
        \centering
        \includegraphics[width=\textwidth]{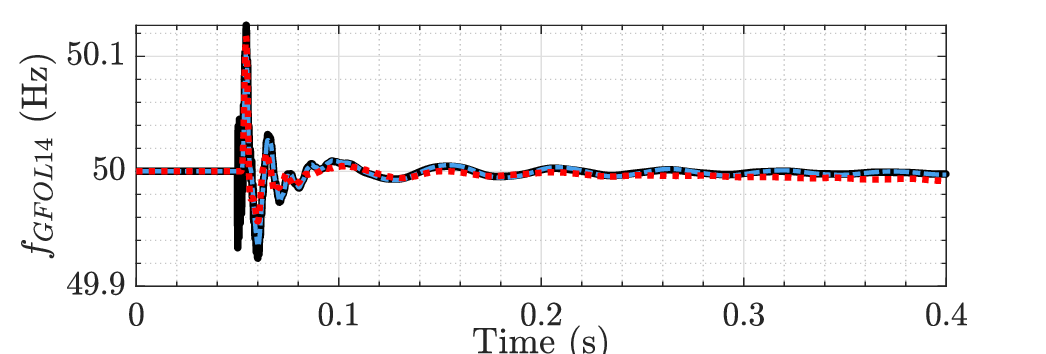}
        \vspace{-6mm}
        \caption{Frequency at bus 14.}
        \label{fig:fGFM}
    \end{subfigure}
    \begin{subfigure}[b]{0.75\columnwidth}
        \centering
        \includegraphics[width=\textwidth]{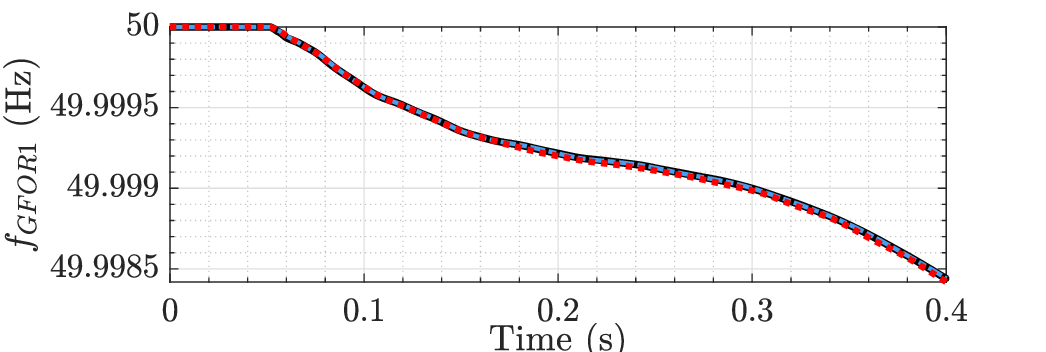}
        \vspace{-6mm}
        \caption{Frequency at bus 1.}
        \label{fig:gfl_f}
        \vspace{-2mm}
    \end{subfigure}
    \caption{SSA-FITSS validation for the New England network.}
    \label{fig:tdvalidation}
    \vspace{-7mm}
\end{figure}

Validation of the extracted SSA-FITSS models was performed by integrating them into the automated STAMP framework~\cite{stamp_citcea} and comparing the results with the SSA-EMT theoretical model and the nonlinear EMT simulation. A 1\% step perturbation in current was applied at bus~1 and time‑domain responses were compared across multiple nodes. As shown in Fig.~\ref{fig:tdvalidation}, the full SSA-FITSS-based model closely matches the the SSA-EMT theoretical and nonlinear EMT responses and successfully captures the dominant dynamics, despite more than half of the VSCs being modeled as black‑box units. Minor discrepancies appear in the initial oscillatory behaviour of voltage, current and frequency signals. As validated in Subsection~\ref{sec:modaldecomposition}, these deviations are attributable to modes not identified during fitting; such modes are neither dominant nor observable from the perturbation location and therefore exert negligible dynamic influence.

\begin{figure}[t!]
\centering
\includegraphics[width=0.8\columnwidth]{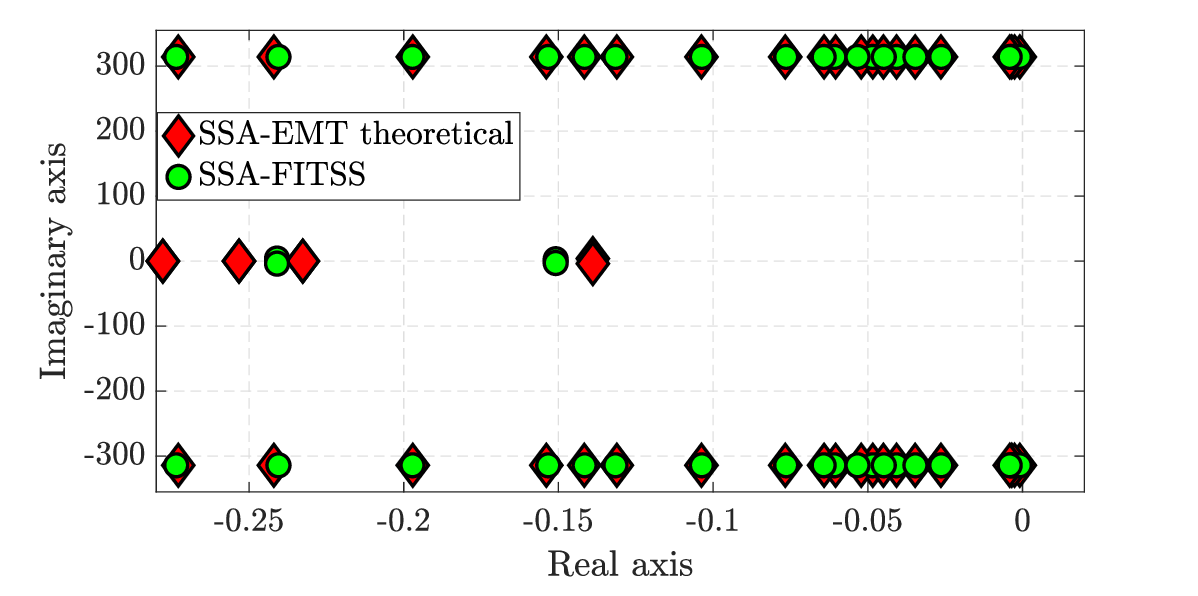}
\vspace{-3.5mm}
\caption{Critical modes from the New England network.}
\label{poles_NewEngland}
\vspace{-2.8mm}
\end{figure}

Fig.~\ref{poles_NewEngland} shows the critical modes of the full SSA‑FITSS‑based model and the SSA‑EMT theoretical model, confirming overall system stability and good agreement with the linearized model. Table~\ref{tab:critical_modes} lists the critical modes from the full SSA‑FITSS-based model, ordered by the smallest damping ratio~$\zeta$, and reports their frequencies and white‑ or black‑box classification using the procedure in \eqref{eq:ss_general}--\eqref{eq:PF_full}. The first ten critical modes are purely white‑box dominated, with damping below 0.001\% and frequencies equal to the nominal system frequency. Purely black‑box dominated modes (from the individual SSA‑FITSS models) are not identified as critical, but they align well with the SSA-EMT theoretical modes and do not introduce redundant or additional critical dynamics.

\begin{table}[t!]
\centering
\caption{Summary of the critical modes.}
\vspace{-1mm}
\label{tab:critical_modes}
\resizebox{\columnwidth}{!}{
\begin{tabular}{cccccc}
\toprule
\textbf{Mode} & \textbf{ID} &  \textbf{Value (rad/s)} & \textbf{$\mathbf{\zeta}$ (\%)} & \textbf{f (Hz)} & Dominant \\
\midrule
642, 643  & 1, 2 & - 0.0008 $\pm$ j314.16  & 0.000 & 50.0 & $wb$ \\ 
640, 641  & 3, 4 & - 0.0025 $\pm$ j314.16  & 0.001 & 50.0 & $wb$ \\
589, 590  & 5, 6 & - 0.0034 $\pm$ j314.16  & 0.001 & 50.0 & $wb$ \\
644, 645  & 7, 8 & - 0.0041 $\pm$ j314.16  & 0.001 & 50.0 & $wb$ \\
591, 592  & 9, 10 & - 0.0263 $\pm$ j314.16  & 0.008 & 50.0 & $wb$ \\
\bottomrule
\end{tabular}
}
\vspace{-2mm}
\end{table}

\begin{figure}[t!]
\centering
\includegraphics[width=0.75\columnwidth]{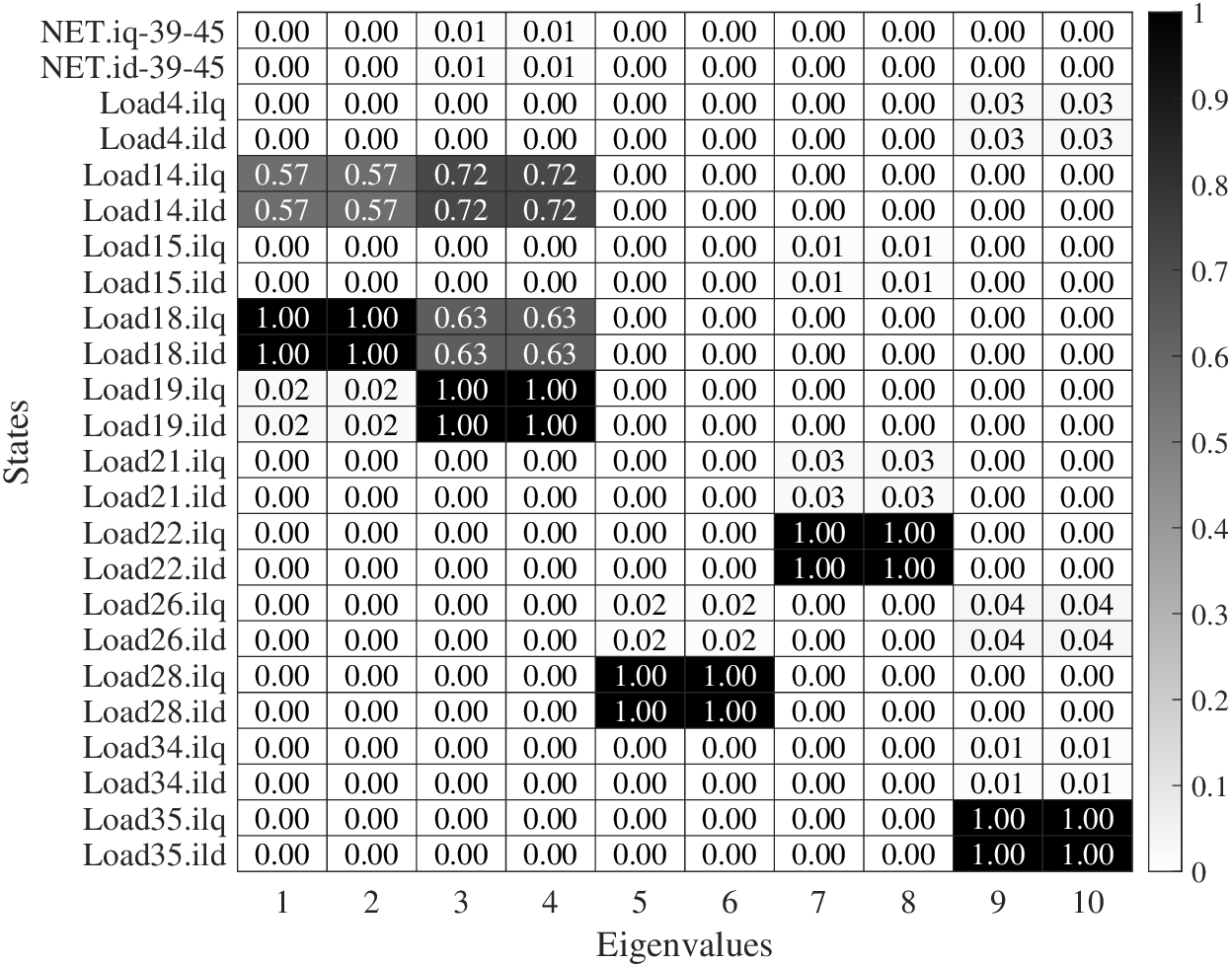}
\vspace{-2mm}
\caption{Summary of the participation factor results.}
\label{fig:PFs}
\vspace{-6mm}
\end{figure}

Finally, a participation factor (PF) analysis is performed on the critical modes of the full SSA‑FITSS model. For clarity, Fig.~\ref{fig:PFs} shows a reduced set of relevant states using a filtering threshold of $PFs>0.005$. The PF results indicate that critical modes 1–2 are dominated by electrical/system variables, with the largest participation from the currents at load~18 and, to a lesser extent, load~14. Modes 3–4 are dominated by load currents at buses 19, 18, and 14. Modes 5–6, 7–8, and 9–10 show similar behavior, dominated by current at loads 28, 22, and 35, respectively. Note that, assigned states from the SSA‑FITSS models do not participate in the critical modes for this case study. These results demonstrate that the combined SSA-FITSS and STAMP framework~\cite{stamp_citcea} provides full access to system-wide modal information, enabling detailed stability assessment in black-box dominated systems.

\vspace{-0.5cm}
\subsection{Where Is the VSC Penetration Limit?}

\subsubsection{SG Reduction}

\begin{figure}[t!]
\centering
\includegraphics[width=0.98\columnwidth]{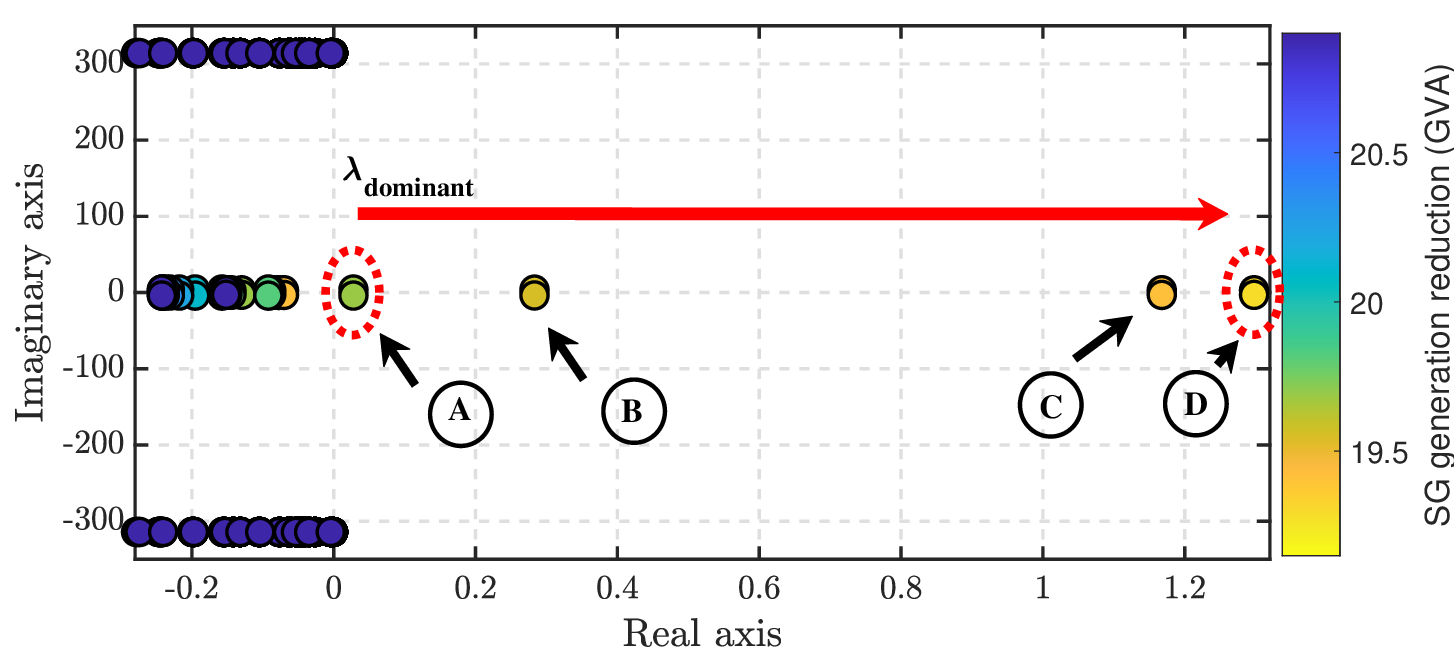}
\vspace{-3mm}
\caption{SSA-FITSS eigenvalue sensitivity for SG reduction.}
\label{fig:polesSG}
\vspace{-6mm}
\end{figure}

In this scenario, the rated capacity of the synchronous generators is progressively reduced to emulate increasing converter penetration. For each operating point, individual SSA-FITSS VSC models are updated and integrated into the full SSA-FITSS-based system. Fig.~\ref{fig:polesSG} shows the critical eigenvalue trajectories as the total SG capacity is reduced from 20\,900~MVA to 19\,150~MVA (an 8.4\% reduction), while the total VSC-based capacity remains 24\,450~MVA. This corresponds to 53.9\% of the total system capacity, with 81.6\% of the VSCs operating in GFL mode and the remaining 18.4\% in GFM mode, representing a VSC‑dominated system.

The sensitivity analysis shows that dominant eigenvalues cross into the right-half plane when SG capacity falls below 19\,700~MVA, revealing system instability after about a 6\% reduction in synchronous generation (confirmed by EMT simulations). The initial unstable modes (modes A in Fig.~\ref{fig:polesSG}) oscillate at 0.64~Hz with a damping ratio of 
$-0.007$ and exhibit the highest participation in the field and stator currents of SG1 and SG2, as confirmed by the participation factor analysis. Modes B at 0.61~Hz with $\zeta=-0.073$ show participation from the same variables. Modes C at 0.49~Hz with $\zeta=-0.352$ are dominated by the same SG‑related variables but primarily from SG1. Modes D, located furthest to the right in the complex plane, again show dominant participation from these variables at 0.48~Hz with $\zeta=-0.394$. This behavior indicates that the instability in the SG‑reduction scenario is mainly driven by SG1 and SG2. Notably, neither the black‑box VSC modes nor their associated variables contribute to the instability.

\subsubsection{GFM Reduction}

\begin{figure}[t!]
\centering
\includegraphics[width=0.985\columnwidth]{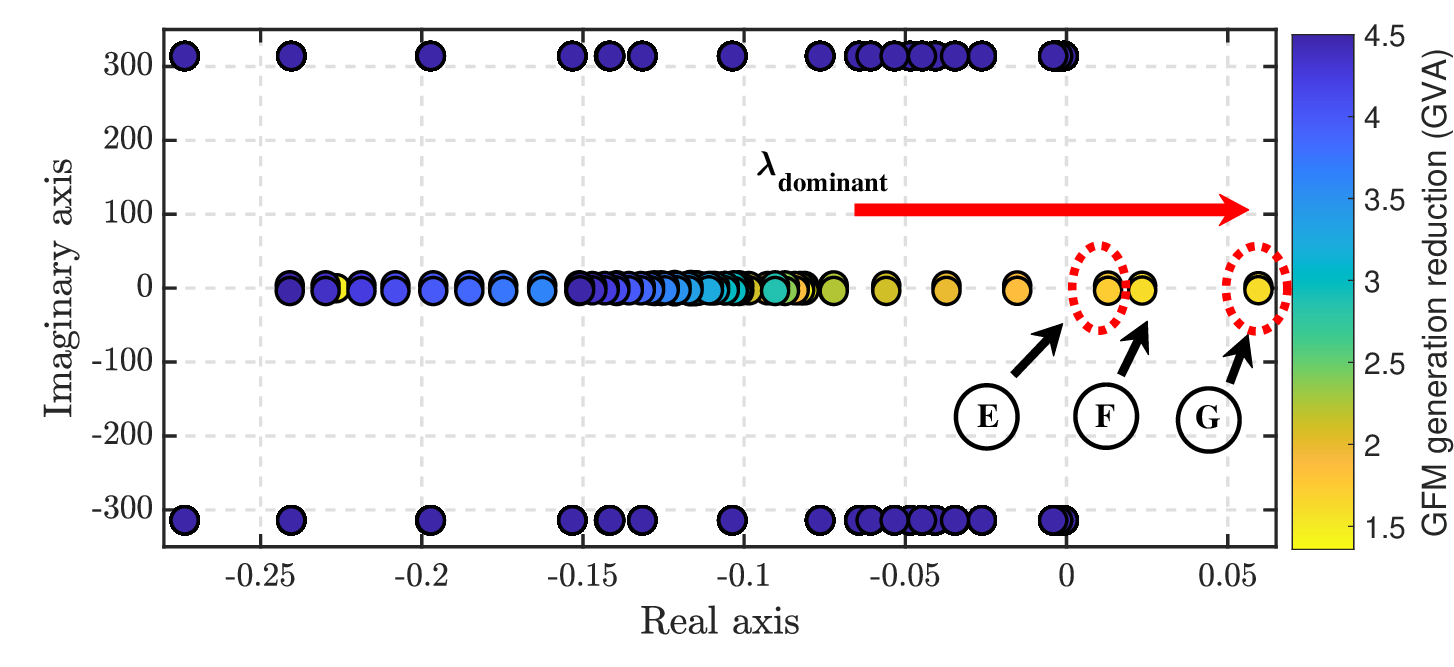}
\vspace{-3mm}
\caption{SSA-FITSS eigenvalue sensitivity for GFM reduction.}
\label{fig:polesGFM}
\vspace{-6.2mm}
\end{figure}

In the second scenario, the number of VSCs operating in GFM mode is reduced, increasing the share of GFL VSCs while maintaining the same SG capacity. The total GFM capacity is initially 4\,500~MVA and is reduced to 1\,360~MVA, corresponding to a 70\% reduction. Fig.~\ref{fig:polesGFM} shows that instability emerges when the GFM capacity falls below 1\,850~MVA, corresponding to 58\% of the initial GFM capacity. 

The participation factor analysis shows that the unstable modes~E at 0.56~Hz with $\zeta=-0.004$ are white-box dominated by the variables {SG2.if-d}, {SG2.is-d}, {SG1.if-d}, and {SG1.is-d}, confirming an electromechanical instability driven by interactions involving the field and stator currents of SG1 and SG2. A minor participation of 0.08 is observed from {GFM3.cc-7} and {GFM3.cc-8}, indicating a very small contribution from the current-control variables of the black-box GFM3 in modes~E. Modes~F at 0.51~Hz with $\zeta=-0.007$ show the largest participation from the frequency and angle of SG4, while still exhibiting contributions from the same SG1 and SG2 variables. For modes~G at 0.43~Hz with $\zeta=-0.022$, the dominant variables remain those of SG1 and SG2, but the participation of the previously mentioned GFM3 variables increases to 0.6, highlighting a growing interaction between the SG units and the black-box GFM3.

Notice that in both scenarios, the SSA-FITSS methodology enables a physically meaningful interpretation of the modes contributing to instability. Although simplified,  these study cases demonstrate the scalability and analytical capability of SSA-FITSS in enabling sensitivity-based small-signal stability assessments in black-box dominated systems.

\vspace{-4mm}
\section{Conclusion}

This work presented an extended multi-variable fitted state-space (SSA-FITSS) methodology for small-signal stability analysis in converter-dominated power systems with limited model data. The method overcomes key limitations of existing identification-based approaches by providing accurate state-space models from black-box VSCs, enabling full eigenvalue-based analysis without requiring access to internal control. The proposed framework integrates frequency-domain identification, adaptive pole-expansion, and model-order reduction into a unified method that scales to large systems. A notable contribution is the automated state-interpretation strategy, which assigns fitted states to representative control-loop categories based on their dominant frequency ranges. This provides an approximate but meaningful physical interpretation of the identified dynamics and enables extensive modal analysis. The New England case study demonstrates that the proposed methodology accurately reproduces converter and system dynamics, identifies stability limits under varying SG and GFM penetration levels, and reveals the dominant mechanisms driving instability. The results confirm that SSA-FITSS enables system-wide stability assessment in scenarios where traditional white-box modeling is not feasible.

\vspace{-0.5cm}
\section*{Acknowledgment}

This work has received funding from the ADOreD project under the European Union’s Horizon Europe Research and Innovation Programme under the Marie Skłodowska-Curie Grant Agreement No. 101073554. The work of O. Gomis-Bellmunt and E. Prieto-Araujo was supported by the Agència de Gestió d’Ajuts Universitaris i de Recerca (AGAUR) through the ICREA Acadèmia programme, and by the Departament de Recerca i Universitats of the Generalitat de Catalunya. The work of E. Prieto‑Araujo was also supported by the Serra Húnter Programme.

\bibliographystyle{IEEEtran}
\vspace{-0.3cm}
\bibliography{ref}
\vfill

\newpage

 




\vfill

\end{document}